\documentstyle{article}
\begin{document}
\title{Hawking's chronology protection conjecture:\\
       singularity structure of the quantum stress--energy tensor}
\author{Matt Visser\cite{e-mail}\\
        Physics Department\\
        Washington University\\
        St. Louis\\
        Missouri 63130-4899}
\date{November 1992}
\maketitle
\begin{abstract}

The recent renaissance of wormhole physics has led to a very
disturbing observation: If traversable wormholes exist then it
appears to be rather easy to to transform such wormholes into time
machines. This extremely disturbing state of affairs has lead
Hawking to promulgate his chronology protection conjecture.

This paper continues a program begun in an earlier paper [Physical
Review {\bf D47}, 554--565 (1993)]. An explicit calculation
of the  vacuum expectation value of the renormalized stress--energy
tensor in wormhole spacetimes is presented. Point--splitting
techniques are utilized. Particular attention is paid to computation
of the Green function [in its Hadamard form],  and the structural
form of the stress-energy tensor near short closed spacelike
geodesics. Detailed comparisons with previous calculations are
presented, leading to a pleasingly unified overview of the situation.


\end{abstract}

\newpage
\section{INTRODUCTION}

This paper addresses Hawking's chronology protection conjecture
\cite{Hawking-I,Hawking-II}, and some of the recent controversy
surrounding this conjecture \cite{Kim-Thorne,Klinkhammer}. The
qualitative and quantitative analyses of reference \cite{CPC} are
extended. Detailed comparisons are made with the analyses of Frolov
\cite{Frolov}, of Kim and Thorne \cite{Kim-Thorne}, and of Klinkhammer
\cite{Klinkhammer}.

The recent renaissance of wormhole physics has led to a very
disturbing observation: If traversable wormholes exist then it
appears to be rather easy to to transform such wormholes into time
machines. This extremely disturbing state of affairs has lead
Hawking to promulgate his chronology protection conjecture
\cite{Hawking-I,Hawking-II}.

This paper continues a program begun in an earlier paper \cite{CPC}.
In particular, this paper will push the analysis of that paper
beyond the Casimir approximation.  To this end, an explicit
calculation of the vacuum expectation value of the renormalized
stress--energy tensor in wormhole spacetimes is presented.
Point--splitting regularization and renormalization techniques are
employed. Particular attention is paid to computation of the Green
function [in its Hadamard form], and the structural form of the
stress-energy tensor near short closed spacelike geodesics.

The computation is similar in spirit to the analysis of Klinkhammer
\cite{Klinkhammer}, but the result appears --- at first blush ---
to be radically different. The reasons for this apparent difference
(and potential source of great confusion) are tracked down and
examined in some detail. Ultimately the various results are shown
to agree with one another in those regions where the analyses
overlap. Furthermore, the  analysis of this paper is somewhat more
general in scope and requires fewer technical assumptions.

\underbar{Notation:} Adopt units where $c\equiv 1$, but all other
quantities retain their usual dimensionalities, so that in particular
$G=\hbar/m_P^2 = \ell_P^2/\hbar$. The metric signature is taken to
be $(-,+,+,+)$.

\section{THE GREEN FUNCTION}

\subsection{The Geodetic Interval}

The geodetic interval  is defined by:
\begin{equation}
\sigma_\gamma(x,y) \equiv \pm{1\over2} [s_\gamma(x,y)]^2.
\end{equation}
Here we take the upper ($+$) sign if the geodesic $\gamma$ from
the point $x$ to the point $y$ is spacelike.  We take the lower
($-$) sign if this geodesic $\gamma$ is timelike. In either case
we define the arc length $s_\gamma(x,y)$ to be positive semi-definite.
Note that, provided the geodesic from $x$ to $y$ is not lightlike,
\begin{eqnarray}
\nabla^x_\mu \sigma_\gamma(x,y)
&=& \pm s_\gamma(x,y) \; \nabla^x_\mu s_\gamma(x,y), \\
&=& + s_\gamma(x,y) \; t_\mu(x;\gamma; x \leftarrow y).
\end{eqnarray}
Here $t_\mu(x; \gamma; x \leftarrow y) \equiv \pm \nabla^x_\mu
s_\gamma(x,y)$ denotes the unit tangent vector at the point
$x$ pointing along the geodesic $\gamma$ away from the point $y$.
When no confusion results we may abbreviate this by $t_\mu(x
\leftarrow y)$ or even $t_\mu(x)$. If the geodesic from $x$ to $y$
is lightlike things are somewhat messier. One easily sees that for
lightlike geodesics $\nabla^x_\mu \sigma_\gamma(x,y)$ is a null
vector. To proceed further one must introduce a canonical observer,
characterized by a unit timelike vector $V^\mu$ at the point $x$.  By
parallel transporting this canonical observer along the geodesic
one can set up a canonical frame that picks out a particular
canonical affine parameter:
\begin{equation}
\nabla^x_\mu \sigma_\gamma(x,y) = +\zeta \; l_\mu; \qquad
l_\mu V^\mu = -1; \qquad
\zeta = -V^\mu \nabla^x_\mu \; \sigma_\gamma(x,y).
\end{equation}
Note that this affine parameter $\zeta$ can, crudely, be thought
of as a distance along the null geodesic as measured by an observer
with four--velocity $V^\mu$.

To properly place these concepts within the context of the chronology
protection conjecture, see, for example, reference \cite{CPC}.
Consider an arbitrary Lorentzian spacetime of nontrivial homotopy.
Pick an arbitrary base point $x$. Since, by assumption, $\pi_1({\cal
M})$ is nontrivial there certainly exist closed paths not homotopic
to the identity that begin and end at $x$. By smoothness arguments
there also exist smooth closed geodesics, not homotopic to the
identity, that connect the point $x$ to itself. However, there is
no guarantee that the tangent vector is continuous as the geodesic
passes through the point $x$ where it is pinned down.  If any of
these closed geodesics is timelike or null then the battle against
time travel is already lost, the spacetime is diseased, and it
should be dropped from consideration.

To examine the types of pathology that arise as one gets ``close''
to building a time machine, it is instructive to construct a one
parameter family of closed geodesics that captures the essential
elements of the geometry.  Suppose merely that one can find a well
defined throat for one's Lorentzian wormhole. Consider the world
line swept out by a point located  in the middle of the wormhole
throat. At each point on this world line there exists a closed
``pinned'' geodesic threading the wormhole and closing back on
itself in ``normal'' space.  This geodesic will be smooth everywhere
except possibly at the place that it is ``pinned'' down by the
throat.

If the geodetic interval from $x$ to itself, $\sigma_\gamma(x,x)$,
becomes negative then a closed timelike curve ({\sl a fortiori}
--- a time machine) has formed.  It is this unfortunate happenstance
that Hawking's chronology protection conjecture is hoped to prevent.
For the purposes of this paper it will be sufficient to consider
the behaviour of the vacuum expectation value of the renormalized
stress energy tensor in the limit $\sigma_\gamma(x,x) \to 0^+$.

\subsection{The Hadamard Form}

The Hadamard form of the Green function may be derived by an
appropriate use of adiabatic techniques
\cite{deWitt-I,deWitt-II,deWitt-III,deWitt-IV,Birrell-Davies,Fulling}:
\begin{eqnarray}
G(x,y)
&\equiv& <0|\phi(x) \phi(y)|0> \nonumber\\
&=& \sum_\gamma {\Delta_\gamma(x,y)^{1/2}\over4\pi^2}
\left[ {1\over\sigma_\gamma(x,y)}
       + \upsilon_\gamma(x,y) \ln|\sigma_\gamma(x,y)|
       + \varpi_\gamma(x,y) \right]_.
\end{eqnarray}
Here the summation, $\sum_\gamma$, runs over all distinct geodesics
connecting the point $x$ to the point $y$. The symbol $\Delta_\gamma(x,y)$
denotes the van Vleck determinant \cite{vanVleck,Morette}.  The
functions $\upsilon_\gamma(x,y)$ and $\varpi_\gamma(x,y)$ are known
to be smooth as $s_\gamma(x,y) \to 0$
\cite{deWitt-I,deWitt-II,deWitt-III,deWitt-IV,Birrell-Davies,Fulling}.

If the points $x$ and $y$ are such that one is sitting on top of
a bifurcation of geodesics (that is: if the points $x$ and $y$ are
almost conjugate) then the Hadamard form  must be modified by the
use of Airy function techniques in a manner similar to that used
when encountering a point of inflection while using steepest descent
methods \cite{deWitt-II}.

\section{RENORMALIZED STRESS--ENERGY}
\subsection{Point splitting}

The basic idea behind point splitting techniques
\cite{deWitt-I,deWitt-II,deWitt-III,deWitt-IV,Birrell-Davies,Fulling}
is to define formally infinite objects in terms of a suitable
limiting process such as
\begin{equation}
<0|T_{\mu\nu}(x)|0> = \lim_{y\to x} <0|T_{\mu\nu}(x,y)|0>.
\end{equation}
The point--split stress--energy tensor $T_{\mu\nu}(x,y)$ is a
symmetric tensor at the point $x$ and a scalar at the point $y$.
The contribution to the point--split stress--energy tensor associated
with a particular quantum field is generically calculable in terms
of covariant derivatives of the Green function of that quantum
field. Schematically
\begin{equation}
<0|T_{\mu\nu}(x,y)|0> = D_{\mu\nu}(x,y) \{G_{\hbox{\rm ren}}(x,y)\}.
\end{equation}
Here $D_{\mu\nu}(x,y)$ is a second order differential operator,
built out of covariant derivatives at $x$ and $y$. The covariant
derivatives at $y$ must be parallel propagated back to the point
$x$ so as to ensure that $D_{\mu\nu}(x,y)$ defines a proper
geometrical object. This parallel propagation requires the introduction
of the notion of the trivial geodesic from $x$ to $y$, denoted by
$\gamma_0$. (These and subsequent comments serve to tighten up,
justify, and make explicit the otherwise rather heuristic incantations
common in the literature.)

Renormalization of the Green function consists of removing the
short distance singularities associated with the flat space Minkowski
limit. To this end, consider a scalar quantum field $\phi(x)$.
Again, let $\gamma_0$ denote the trivial geodesic from $x$ to $y$.
Define
\begin{equation}
G_{\hbox{\rm ren}}(x,y)
\equiv G(x,y) - {\Delta_{\gamma_0}(x,y)^{1/2}\over4\pi^2}
\left[ {1\over\sigma_{\gamma_0}(x,y)}
       + \upsilon_{\gamma_0}(x,y) \ln|\sigma_{\gamma_0}(x,y)| \right]_.
\end{equation}
These subtractions correspond to a wave--function renormalization
and a mass renormalization respectively. Note that any such
renormalization prescription is always ambiguous up to further
finite renormalizations. The scheme described above may profitably
be viewed as a modified minimal subtraction scheme.  In particular,
(for a free quantum field in a curved spacetime), this renormalization
prescription is sufficient to render $<0|\phi^2(x)|0>$ finite:
\begin{eqnarray}
<0|\phi^2(x)|0>_{\hbox{\rm ren}}
=&& G_{\hbox{\rm ren}}(x,x), \nonumber \\
=&& {\Delta_{\gamma_0}(x,x)^{1/2} \varpi_{\gamma_0}(x,x)\over4\pi^2}
    \nonumber\\
 && + \mathop{{\sum}'}_\gamma {\Delta_\gamma(x,x)^{1/2}\over4\pi^2}
       \times \nonumber\\
 && \qquad   \left[ {1\over\sigma_\gamma(x,x)}
           + \upsilon_\gamma(x,x) \ln|\sigma_\gamma(x,x)|
           + \varpi_\gamma(x,x) \right]_.
\end{eqnarray}

For the particular case of a conformally coupled massless scalar
field \cite{Klinkhammer}:
\begin{eqnarray}
D_{\mu\nu}(x,y) &\equiv&
{1\over6}\left( \nabla^x_\mu \; g_\nu{}^\alpha(x,y)\nabla^y_\alpha
              + g_\mu{}^\alpha(x,y) \nabla^y_\alpha \nabla^x_\nu \right)
\nonumber\\
&-&{1\over12} g_{\mu\nu}(x) \left(
        g^{\alpha\beta}(x,y) \nabla^x_\alpha \nabla^y_\beta \right)
\nonumber\\
&-&{1\over12}\left(\nabla^x_\mu \nabla^x_\nu
                  + g_\mu{}^\alpha(x,y) \nabla^y_\alpha \;
                    g_\nu{}^\beta(x,y)  \nabla^y_\beta \right)
\nonumber\\
&+&{1\over48} g_{\mu\nu}(x)
        \left( g^{\alpha\beta}(x) \nabla^x_\alpha \nabla^x_\beta
             + g^{\alpha\beta}(y) \nabla^y_\alpha \nabla^y_\beta  \right)
\nonumber\\
&-&R_{\mu\nu}(x) + {1\over4} g_{\mu\nu}(x) R(x).
\end{eqnarray}
As required, this object is a symmetric tensor at $x$, and is a
scalar at $y$.  The bi-vector $g_\mu{}^\nu(x,y)$ parallel propagates
a vector at $y$ to a vector at $x$, the parallel propagation being
taken along the trivial geodesic $\gamma_0$. (The effects of this
parallel propagation can often be safely ignored, {\sl vide}
reference \cite{Klinkhammer} equation (3), and reference \cite{Frolov}
equation (2.40).)

\subsection{Singularity structure}

To calculate the renormalized stress--energy tensor one merely
inserts the \newline Hadamard form of the Green function (propagator) into
the point split formalism \cite{Kim-Thorne,Klinkhammer,Frolov}.
\begin{equation}
<0|T_{\mu\nu}(x)|0>  =
\mathop{{\sum}'}_\gamma
     {\Delta_\gamma(x,x)^{1/2}\over4\pi^2} \lim_{y\to x}
     D_{\mu\nu}(x,y) \left\{ {1\over\sigma_\gamma(x,y)} \right\}
           + O(\sigma_\gamma(x,x)^{-3/2}).
\end{equation}
Observe that
\begin{equation}
\nabla^x_\mu \; \nabla^y_\nu \left\{ {1\over\sigma_\gamma(x,y)} \right\}
= {1\over\sigma_\gamma(x,y)^2}
\left\{ 4 \nabla^x_\mu s_\gamma(x,y)\;
          \nabla^y_\nu s_\gamma(x,y) -
	  \nabla^x_\mu \; \nabla^y_\nu \sigma_\gamma(x,y) \right\}.
\end{equation}
Similar equations hold for other combinations of derivatives.
Let $y\to x$, and define $t^x_\mu \to t^1_\mu$; $t^y_\mu \to
t^2_\mu$. Note that $g_{\alpha\beta}(x) t^x_\alpha  t^x_\beta =
+1$, since the self--connecting geodesics are all taken to be
spacelike. Then, keeping only the most singular term
\begin{equation}
<0|T_{\mu\nu}(x)|0>  =
\mathop{{\sum}'}_\gamma
     {\Delta_\gamma(x,x)^{1/2}\over4\pi^2\sigma_\gamma(x,x)^2}
      \; \left( t_{\mu\nu}(x;\gamma) + s_{\mu\nu}(x;\gamma) \right)
      + O(\sigma_\gamma(x,x)^{-3/2}).
\label{stress-tensor-I}
\end{equation}
Here the dimensionless tensor $t_{\mu\nu}(x;\gamma)$ is constructed
solely out of the metric and the tangent vectors to the geodesic
$\gamma$ as follows
\begin{equation}
t_{\mu\nu}(x;\gamma) =
{2\over3}\left( t^1_\mu t^2_\nu
              + t^2_\mu t^1_\nu
	      - {1\over2}  g_{\mu\nu} (t^1\cdot t^2) \right)
-{1\over3}\left(t^1_\mu t^1_\nu
              +  t^2_\mu t^2_\nu  -{1\over2} g_{\mu\nu}  \right)_.
\label{structure}
\end{equation}
The dimensionless tensor $s_{\mu\nu}(x;\gamma)$ is defined by
\begin{equation}
s_{\mu\nu}(x;\gamma) \equiv \lim_{y\to x}
     D_{\mu\nu}(x,y) \left\{ \sigma_\gamma(x,y) \right\}.
\end{equation}
In many cases of physical interest the tensor $s_{\mu\nu}(x;\gamma)$
either vanishes identically or is subdominant in comparison to
$t_{\mu\nu}(x;\gamma)$. A general analysis has so far unfortunately
proved elusive. This is an issue of some delicacy that clearly
needs further clarification. Nevertheless the neglect of
$s_{\mu\nu}(x;\gamma)$ in comparison to $t_{\mu\nu}(x;\gamma)$
appears to be a safe approximation which shall be adopted forthwith.

Note that this most singular contribution to the stress energy
tensor is in fact traceless --- there is a good physical reason
for this.  Once the length of the closed spacelike geodesic becomes
smaller than the Compton wavelength of the particle under consideration,
$s << \hbar/mc$, one expects such a physical particle to behave in
an effectively massless fashion. Indeed, based on such general
considerations, one expects the singular part of the stress--energy
tensor to be largely insensitive to the type of particle under
consideration. Despite the fact that the calculation has been
carried out only for conformally coupled massless scalars, one
expects this leading singularity to be generic.  Indeed, in terms
of the geodesic distance from $x$ to itself:
\begin{equation}
<0|T_{\mu\nu}(x)|0>
= \mathop{{\sum}'}_\gamma
  {\Delta_\gamma(x,x)^{1/2}\over\pi^2 s_\gamma(x,x)^4}
   \; t_{\mu\nu}(x;\gamma) + O(s_\gamma(x,x)^{-3}).
\label{stress-tensor-II}
\end{equation}
A formally similar result was obtained by Frolov in reference
\cite{Frolov}. That result was obtained for points near the $N$'th
polarized hypersurface of a locally static spacetime. It is important
to observe that in the present context it has not proved necessary
to introduce any (global or local) static restriction on the
spacetime.  Neither is it necessary to introduce the notion of a
polarized hypersurface. All that is needed at this stage is the
existence of at least one short, nontrivial, closed, spacelike
geodesic.

To convince oneself that the apparent $s^{-4}$ divergence of the
renormalized stress--energy is neither a coordinate artifact nor
a Lorentz frame artifact consider the scalar invariant
\begin{equation}
{\cal T} = \sqrt{ <0|T_{\mu\nu}(x)|0> \; <0|T^{\mu\nu}(x)|0> }.
\end{equation}
By noting that
\begin{equation}
t_{\mu\nu}(x;\gamma) t^{\mu\nu}(x;\gamma) =
{1\over3}\left[3 - 4(t^1\cdot t^2) + 2 (t^1\cdot t^2)^2 \right]
\end{equation}
one sees that there is no ``accidental'' zero in $t^{\mu\nu}$, and
that ${\cal T}$ does in fact diverge as $s^{-4}$.  Thus the $s^{-4}$
divergence encountered in the stress--energy tensor associated with
the Casimir effect \cite{CPC} is generic to any multiply connected
spacetime containing short closed spacelike geodesics.

\subsection{Wormhole disruption}

To get a feel for how this divergence in the vacuum polarization
back reacts on the geometry, recall, following Morris and Thorne
\cite{Morris-Thorne,MTY} that a traversable wormhole must be threaded
by some exotic stress energy to prevent the throat from collapsing.
In particular, at the throat itself (working in Schwarzschild
coordinates) the total stress--energy tensor takes the form
\begin{equation}
T_{\mu\nu} = {\hbar\over \ell_P^2 R^2} \left[
\begin{array}{cccc}
\xi\quad&0\quad  &0\quad&0     \\
0       &\chi    &0     &0     \\
0       &0       &\chi  &0     \\
0       &0       &0     &-1
\end{array}
\right]
\end{equation}
On general grounds $\xi <1$, while $\chi$ is unconstrained. In
particular, to prevent collapse of the wormhole throat, the scalar
invariant ${\cal T}$ must satisfy
\begin{equation}
{\cal T}={\hbar\over\ell_P^2 R^2}\sqrt{1+\xi^2+2\chi^2}
\approx {\hbar\over\ell_P^2 R^2}.
\end{equation}

On the other hand, consider the geodesic that starts at a point on
the throat and circles round to itself passing through the throat
of the wormhole exactly once. That geodesic, by itself, contributes
to the vacuum polarization effects just considered an amount
\begin{equation}
<0|T_{\mu\nu}(x)|0>
= {\Delta_\gamma(x,x)^{1/2}\over\pi^2 s_\gamma(x,x)^4}
   \; t_{\mu\nu}(x;\gamma) + O(s_\gamma(x,x)^{-3}).
\end{equation}
So the contribution of the single pass geodesic to the invariant
${\cal T}$ is already
\begin{equation}
{\cal T}
= {\Delta_\gamma(x,x)^{1/2}\over\pi^2 s_\gamma(x,x)^4}
  \sqrt{1-{4\over3}(t^1\cdot t^2) + {2\over3}(t^1\cdot t^2)^2}
  \approx {\Delta_\gamma(x,x)^{1/2}\over\pi^2 s_\gamma(x,x)^4}.
\end{equation}

Therefore, provided that there is no accidental zero in the van
Vleck determinant, vacuum polarization effects dominate over the
wormhole's internal structure once
\begin{equation}
s_\gamma(x,x)^2
<< \ell_P R.
\end{equation}
Indeed, Kim and Thorne have argued \cite{Kim-Thorne} as follows:
In the geometry presently under consideration, (a point $x$ on the
wormhole throat, a geodesic $\gamma$ that loops once around the
wormhole), the thin wall approximation for the throat of the wormhole
leads to a van Vleck determinant equal to unity: $\Delta_\gamma(x,x)=1$.
That derivation is subordinate to a particular choice of identification
scheme for the wormhole mouths, ``time--shift identification'';
but the result holds also for ``synchronous identification''
\cite{CPC}.

More generally, this discussion serves to focus attention on the
van Vleck determinant.  Relatively little is known about the
behaviour of the van Vleck determinant for arbitrary geometries
--- and this is clearly a subject of considerable mathematical and
physical interest.  In particular, a corollary of the previous
comments is that if one could show that a zero of the van Vleck
determinant could be made to coincide with the onset of time machine
formation then one would have strong evidence that singularities
in the quantum stress--energy tensor are not a sufficiently strong
physical mechanism to enforce the chronology protection conjecture.

\section{RELATIONSHIP TO PREVIOUS WORK}

The characteristic $s^{-4}$ divergence encountered in this and
previous analyses \cite{CPC} is, at first blush, somewhat difficult
to reconcile with the ``$(\delta t)^{-3}$'' behaviour described in
references \cite{Kim-Thorne,Klinkhammer,Frolov}. These apparent
differences are, for the most part, merely artifacts due to an
unfortunate choice of Lorentz frame. To see how this happens, one
first has to add considerably more structure to the discussion in
the form of extra assumptions.

To begin the comparison, one must beg the original question by
assuming that a time machine does in fact succeed in forming.
Further, one must assume that the resulting chronology horizon is
compactly generated \cite{Hawking-I,Hawking-II,Kim-Thorne}. The
generators of the compactly generated chronology horizon all converge
in the past on a unique closed null geodesic that shall be referred
to as the ``fountain'', and shall be denoted by $\tilde\gamma$.
The question of interest is now the behaviour of the renormalized
stress energy tensor in the neighborhood of the fountain.

To that end, pick a point $x$ ``close'' to the fountain $\tilde\gamma$.
Pick a point $x_0$ that is on the fountain, with $x_0$ being
``close'' to $x$, and with $x_0$ being in the future of $x$. Then
the geodesic $\gamma_\perp$ from $x$ to $x_0$ is by construction
timelike. One defines $\delta t = s_{\gamma_\perp}(x,x_0)$, and
$V^\mu = -\nabla_x^\mu \sigma_{\gamma_\perp}(x,x_0)$. One interprets
these definitions as follows: a geodesic observer at the point $x$,
with four-velocity $V^\mu$, will hit the fountain $\tilde\gamma$
after a proper time $\delta t$ has elapsed.  One now seeks a
computation of the stress--energy tensor at $x$ in terms of various
quantities that are  Taylor series expanded  around the assumed
impact point $x_0$ with $\delta t$ as the (hopefully) small parameter.

Consider, initially, the geodetic interval $\sigma_\gamma(x,y)$.
Taylor series expand this as
\begin{equation}
\sigma_\gamma(x,x)
= \sigma_{\tilde\gamma}(x_0,x_0)
+ (\delta t\; V^\mu)
\left.
\left[  \nabla^x_\mu \sigma_{\tilde\gamma}(x,y)
      + \nabla^y_\mu \sigma_{\tilde\gamma}(x,y) \right]
\right|_{(x_0,x_0)}
+ O(\delta t^2).
\end{equation}
Firstly, by definition of the fountain as a closed null geodesic,
$\sigma_{\tilde\gamma}(x_0,x_0) = 0$. Secondly, take the vector
$V^\mu$, defined at the point $x$, and parallel propagate it along
$\gamma_\perp$ to $x_0$. Then parallel propagate it along the
fountain $\tilde\gamma$. This now gives us a canonical choice of affine
parameter on the fountain. Naturally this canonical affine parameter
is not unique, but depends on our original choice of $V^\mu$ at
$x$, or, what amounts to the same thing, depends on our choice of
$x_0$ as a ``reference point''. In terms of this canonical affine
parameter, one sees
\begin{eqnarray}
\left. \nabla^x_\mu \sigma_{\tilde\gamma}(x,y) \right|_{(x_0,x_0)} &=&
-\zeta_n^\leftarrow \; l_\mu; \\
\left. \nabla^y_\mu \sigma_{\tilde\gamma}(x,y) \right|_{(x_0,x_0)} &=&
-\zeta_n^\rightarrow \; l_\mu.
\end{eqnarray}
Here the notation $\zeta_n^\leftarrow$ denotes the lapse of affine
parameter on going around the fountain a total of $n$ times in the
left direction, while $\zeta_n^\rightarrow$ is the lapse of affine
parameter for $n$ trips in the right direction. The fact that these
total lapses are different is a reflection of the fact that the
tangent vector to the fountain undergoes a boost on travelling
round the fountain. Hawking showed that \cite{Hawking-I,Hawking-II}
\begin{equation}
\zeta_n^\leftarrow = - e^{nh} \zeta_n^\rightarrow.
\end{equation}
So, dropping explicit exhibition of the $\leftarrow$,
\begin{equation}
\sigma_\gamma(x,x)
=   + \delta t  \; \zeta_n \left[ e^{nh} -1 \right]
+ O(\delta t^2).
\end{equation}
This, finally, is the precise justification for equation (5) of
reference \cite{Klinkhammer}.

In an analogous manner, one estimates the tangent vectors $t^1$
and $t^2$ in terms of the tangent vector at $x_0$:
\begin{eqnarray}
\left.\nabla^x_\mu \sigma_\gamma(x,y)\right|_{y\to x}
&=&
\left.\nabla^x_\mu \sigma_{\tilde\gamma}(x,y)\right|_{(x_0,x_0)}
    + O(\delta t)\\
&=& -\zeta_n l_\mu + O(\delta t).
\end{eqnarray}
This leads to the estimate
\begin{equation}
t_1^\mu \approx - e^{nh} \; t_2^\mu \approx -(\zeta_n/s) \; l_\mu.
\label{tangents}
\end{equation}
\underline{Warning:} This estimate should be thought of as an
approximation for the dominant {\sl components} of the various
vectors involved. If one takes the norm of these vectors one
finds
\begin{equation}
1 \approx  e^{nh} \approx 0.
\end{equation}
This is true in the sense that other components are larger, but
indicates forcefully the potential difficulties in this approach.

One is now ready to tackle the estimation of the structure tensor
$t_{\mu\nu}(x;\gamma)$. Using equations (\ref{structure}) and
(\ref{tangents}) one obtains
\begin{equation}
t_{\mu\nu}(x;\gamma) \approx
- {\zeta_n^2\over3 s^2} \left[1 + 4e^{nh} + e^{2nh} \right]
 l_\mu l_\nu.
\end{equation}

Pulling the various estimates together, the approximation to the
Hadamard stress--energy tensor is seen to be
\begin{equation}
<0|T_{\mu\nu}(x)|0>  =
- \mathop{{\sum}'}_\gamma
\left\{
     {\Delta_\gamma(x,x)^{1/2}\over24\pi^2\zeta_n}
     {\left[1 + 4e^{nh} + e^{2nh} \right] \over
      \left[e^{nh} -1\right]^3}
\right\}
      \; {l_\mu l_\nu \over (\delta t)^3}
      + O(\delta t^{-2}).
\label{KT-form}
\end{equation}
This, finally, is exactly the estimate obtained by Klinkhammer
\cite{Klinkhammer} --- his equation (8). Furthermore this result
is consistent with that of Kim and Thorne \cite{Kim-Thorne} ---
their equation (67). The somewhat detailed presentation of this
derivation has served to illustrate several important points.

Primus, the present result is a special case of the more general
result (\ref{stress-tensor-II}), the present result being obtained
only at the cost of many additional technical assumptions.  The
previous analysis has shown that the singularity structure of the
stress--energy tensor may profitably be analysed without having to
restrict attention to regions near the fountain of a compactly
generated chronology horizon. The existence of at least one short,
closed, nontrivial, spacelike geodesic is a sufficient requirement
for the extraction of useful information.

Secundus, the approximation required to go from (\ref{stress-tensor-II})
to (\ref{KT-form}) are subtle and potentially misleading.  For
instance, calculating the scalar invariant ${\cal T}$ from
(\ref{KT-form}), the leading $(\delta t)^{-3}$ term vanishes (because
$l^\mu$ is a null vector). The potential presence of a subleading
$(\delta t)^{-5/2}$ cross term cannot be ruled out from the present
approximation, (\ref{KT-form}).  Fortunately, we already know [from
the original general analysis, (\ref{stress-tensor-II}) ] that the
dominant behaviour of ${\cal T}$  is ${\cal T}\propto \sigma^{-2}\propto
s^{-4}$.  In view of the fact that, under the present restrictive
assumptions, $\sigma \propto  \delta t$, one sees that ${\cal
T}\propto (\delta t)^{-2}$. The cross term, whatever it is, must
vanish.

Tertius, a warning --- this derivation serves to expose, in
excruciating detail, that the calculations encountered in this
problem are sufficiently subtle that two apparently quite different
results may nevertheless be closely related.

\section{DISCUSSION}

This paper has investigated the leading divergences in the vacuum
expectation value of the renormalized stress--energy tensor as the
geometry of spacetime approaches time machine formation.  Instead
of continuing the ``defense in depth'' strategy of the author's
previous contribution \cite{CPC}, this paper focuses more precisely
on wormhole disruption effects. If one wishes to use a traversable
wormhole to build a time machine, then one must somehow arrange to
keep that wormhole open. However, as the invariant distance around
and through the wormhole shrinks to zero the stress--energy at the
throat diverges. Vacuum polarization effects overwhelm the wormhole's
internal structure once
\begin{equation}
s_\gamma(x,x)^2
<< \ell_P R.
\end{equation}
This happens for $s_\gamma(x,x) >> \ell_P$, which fact I interpret
as supporting Hawking's chronology protection conjecture. This
result is obtained without invoking the technical requirement of
the existence of a compactly generated chronology horizon.  If such
additional technical assumptions are added, the formalism may be
used to reproduce the known results of Kim and Thorne \cite{Kim-Thorne},
of Klinkhammer \cite{Klinkhammer}, and of Frolov \cite{Frolov}.

Moving beyond the immediate focus of this paper, there are still
some technical issues of considerable importance left unresolved.
Most particularly, computations of the van Vleck determinant in
generic traversable wormhole spacetimes is an issue of some interest.

\underline{Acknowledgements:}

I wish to thank Carl Bender and Kip Thorne for useful discussions.
I also wish to thank the Aspen Center for Physics for its hospitality.
This research was supported by the U.S. Department of Energy.


\end{document}